\documentclass[letterpaper]{article} % DO NOT CHANGE THIS
\usepackage{aaai2026}  % DO NOT CHANGE THIS
\usepackage{times}  % DO NOT CHANGE THIS
\usepackage{helvet}  % DO NOT CHANGE THIS
\usepackage{courier}  % DO NOT CHANGE THIS
\usepackage[hyphens]{url}  % DO NOT CHANGE THIS
\usepackage{graphicx} % DO NOT CHANGE THIS
\usepackage{natbib}  % DO NOT CHANGE THIS AND DO NOT ADD ANY OPTIONS TO IT
\usepackage{caption} % DO NOT CHANGE THIS AND DO NOT ADD ANY OPTIONS TO IT
\usepackage{graphicx}
\usepackage{amsmath}

\usepackage{algorithm}
\usepackage[noend]{algorithmic}

\usepackage{xcolor}

\usepackage{multirow}

\usepackage{newfloat}
\usepackage{listings}
\DeclareCaptionStyle{ruled}{labelfont=normalfont,labelsep=colon,strut=off} % DO NOT CHANGE THIS
\floatstyle{ruled}
\newfloat{listing}{tb}{lst}{}
\floatname{listing}{Listing}
\title{Flow-Based Task Assignment for \\ Large-Scale Online Multi-Agent Pickup and Delivery}
\author{
    Yue Zhang,
    Zhe Chen,
    Daniel Harabor, 
    Pierre Le Bodic,
    Peter J. Stuckey
}
\affiliations{
    Monash University, Australia\\
    \{Yue.Zhang, Zhe.Chen, Daniel.Harabor, Pierre.LeBodic, Peter.Stuckey\}@monash.edu
}

\usepackage{bibentry}
\begin{document}

\maketitle
\begin{abstract}
We study the problem of online Multi-Agent Pickup and Delivery (MAPD), where a team of agents must repeatedly serve dynamically appearing tasks on a shared map. 
Existing online methods either rely on simple heuristics, which result in poor decisions, or employ complex reasoning, which suffers from limited scalability under real-time constraints.
In this work, we focus on the task assignment subproblem and formulate it as a minimum-cost flow over the environment graph.
This eliminates the need for pairwise distance computations and allows agents to be simultaneously assigned to tasks and routed toward them.
The resulting flow network also supports efficient guide path extraction to integrate with the planner and accelerates planning under real-time constraints.
To improve solution quality, we introduce two congestion-aware edge cost models that incorporate real-time traffic estimates.
This approach supports real-time execution and scales to over 20000 agents and 30000 tasks within 1-second planning time, outperforming existing baselines in both computational efficiency and assignment quality.
%Some existing MAPD solvers operate offline, assuming there is enough computation budget and all tasks are known in advance. However, this assumption is not consistent with real-world applications, where tasks arrive dynamically and agents must make decisions online during execution.
\end{abstract}


\section{Introduction}
Multi-Agent Pickup and Delivery (MAPD) is a fundamental problem in autonomous multi-robot systems that requires a team of agents to continuously execute a large amount of tasks distributed over a shared environment while avoiding collisions with each other \cite{ma2017lifelong_token_passing}.
%, with tasks either being known in advance or continuously received during the execution.
In MAPD, each task consists of transporting an item from a pickup location to a delivery location, and agents operate in an online setting where tasks arrive continuously and must be assigned and executed in real time. 
This problem has wide applications in warehouse automation, autonomous aircraft-towing vehicles \cite{morris2016planning}, office robots \cite{veloso2015cobots} and video games \cite{ma2017feasibility}.

Solving MAPD involves two subproblems: assigning available tasks to agents and planning collision-free paths for agents to complete these tasks. 
A core challenge in this problem lies in balancing task assignment and path planning under strict runtime constraints, and in the presence of thousands or even tens of thousands of agents and tasks.
A decision must be made at each planning window: which agent should be assigned to which task, and what path should they follow. 
As a result, solvers must react quickly to task arrivals and system state changes, while also avoiding congestion and conflicts during execution. This makes the joint optimisation of task assignment and path planning especially difficult at scale to meet strict time constraints.

To address the online setting, Token Passing (TP) and Token Passing with Task Swaps (TPTS) were proposed \cite{ma2017lifelong_token_passing}. TP assigns tasks in a greedy manner by passing a token among agents, allowing them to claim tasks and plan paths one by one. TPTS further allows task swaps between agents in TP. While conceptually simple and reactive, TP and TPTS often produce suboptimal assignments and suffer from bottlenecks from time-dependent path planning or replanning, especially as the team size increases. 
%(i.e., each agent selects a target and plans a complete path, before passing the token), especially as the team size increases. 
To further optimise the solution quality, RMCA \cite{chen2021integrated}
optimise the solution by integrating and solving them as a combined problem. 
However, these works still suffer from computation bottlenecks when scaling to hundreds of agents.

Similarly, some existing approaches solves the problem offline, where all tasks are known in advance, and try to compute globally optimal solutions for both agent-task assignment and path planning within a given runtime \cite{honig2018conflict_cbs_ta,lam2025optimal_bcp_mapd,liu2019task}.
These algorithms can provide high-quality solutions. However, they assume all tasks are known in advance and enough planning time is given upfront, i.e., one hour. These algorithms can provide high-quality solutions. As a result, they are not designed for online operations, where tasks are not all known \textit{a priori}, and solvers are required to react quickly to avoid agents being left waiting for new tasks. In addition, they also do not scale well beyond dozens of agents due to the high computational complexity of joint reasoning.

This work focuses on the assignment side of MAPD.
Once all-pair distances are computed between available agents and tasks, the assignment is a special case of the minimum-cost flow problem, hence it can be solved rather efficiently using the network simplex \cite{ahuja1993network}, for instance.
However, computing the all-pair distance matrix can, by itself, be prohibitive.

Instead, we propose a new flow-based framework for large-scale task assignment in online MAPD. Our key idea is to solve the assignment without computing distances upfront, but directly on the map, as a minimum-cost flow. 
The resulting flow not only determines the assignments but also produces guide paths for agents, which can be integrated into the path planner to accelerate planning by reducing search overhead, which helps produce high-quality path plans within the tight time constraints.
While our primary contribution is the flow-based model, we further introduce traffic-aware edge cost models that incorporate real-time congestion estimates into the flow network, encouraging the system to avoid congested areas and distribute agents more evenly.
In the experiment, our approach outperforms existing baselines in terms of solution quality and runtime and scales to scenarios with over 20000 agents and 30000 tasks on large maps
Our method runs in real time, offers high throughput, and allows for flexible integration with different cost models and traffic estimators.

\section{Problem Setup}

%\subsection{MAPD}
A \textit{Multi-Agent Pickup and Delivery} (MAPD) problem consists of $n$ agents $A = \{a_1, \ldots, a_n\}$ on a known 2D grid map $G=(V,E)$, where $V$ is a set of vertices (grid cells) and $E$ is a set of edges that connects adjacent cells. 
Each agent starts at a unique location and is responsible for repeatedly completing delivery tasks.
Each task consists of a pair of locations, a pickup location and a delivery location. To complete a task, an agent must reach the pickup location and then move to the delivery location in order. 
%\subsection{Online MAPD} In the online version of MAPD, 
Tasks are not known in advance. Instead, they appear dynamically over time. A global \textit{task pool} maintains all currently available tasks. Let $m$ be the number of tasks available. Once a task is completed, a new task is released into the pool to maintain a constant number of tasks in the pool (or according to a predefined release policy).

The system runs in discrete time steps. At each step, the solver must assign tasks to available agents and update plans for ongoing assignments. Agents can move to adjacent free cells or wait in place. Collision avoidance must be enforced: agents cannot occupy the same cell at the same time or traverse the same edge in opposite directions simultaneously. The goal is to assign tasks and plan paths to free agents in a way that maximises overall throughput (i.e., the number of completed tasks over time)

We adopt an online execution model similar to that used in \citet{zhang2024planning_pie} and \citet{chan2024league}, where solvers plan while executing. In this setting, at each step, the solver is given a fixed planning time window, i.e., determined by the execution time for a single action, to compute both task assignments and movement decisions. If the solver does not return within this time limit, agents will pause and wait during that step, leading to delays in task completion. 

\section{Related Work}
\subsection{Path Planning Approaches}
MAPD consists of two subproblems: task assignment and path planning. The path planning problem is a well-studied problem called Multi-agent path finding (MAPF) \cite{stern2019multi}, where a team of agents navigate from the given start to goal positions while avoiding collisions. In this problem, each agent receives only one task, and the tasks are assumed to be given and fixed in advance. 
Classical approaches include centralised solvers like Conflict-Based Search (CBS)~\cite{sharon2015conflict} and its many variants, which offer good solution quality, but scale poorly with the number of agents. 
More recent work explores online MAPF, where planning and execution are concurrent \cite{zhang2024planning_pie}. In this online setting, solvers must return plans or partial plans within a fixed planning time, called \textit{committed paths}, and agents act on committed paths while planners are planning for future paths. 
The path planning under this setting becomes more challenging, as planners should react quickly. 
Solvers that are able to produce partial solutions quickly become more desirable. For example, \citet{zhang2024planning_pie} proposes to use a fast method to compute an initial solution, then commit the first several actions and keep improving the uncommitted part of the solutions during execution.
Similar ideas have also been proposed in \citet{chen2024traffic_planner}.
%, where the planner computes and optimises a future spatial guide path for each agent during execution based on a future traffic prediction. 
At each step, the solver commits only the next action for each agent, and keeps improving a spatial guide path during execution time. This method scales well to 10000 agents on various maps.

\subsection{Task Assignment Approaches}
In this context, the task assignment problem consists in finding a minimum-cost set of $\min(n,m)$ disjoint edges going from agents to tasks.
This can then be solved optimally using the Hungarian Method \cite{kuhn1955hungarian} or network flow solvers. 
Related problems have also been widely studied in different areas, 
such as the multi-robot task allocation (MRTA), where a team of robots need to visit a set of target locations \cite{korsah2013comprehensive_mrta,gerkey2004formal_mrta2, lagoudakis2005auction_mrta3}, and the vehicle routing problem (VRP) where multiple vehicles need to deliver products to a group of customers \cite{laporte2009fifty_vrp, lenstra1981complexity_vrp2}.
Additional variations of this problem incorporate different constraints, such as 
adding deadlines or precedence constraints for tasks \cite{bai2019efficient} 
and introducing time windows to tasks and robots. \cite{potvin1993parallel}. 
Approaches in these topics focus more on optimising under a complex model and assignment constraints from the problem.
However, they often rely on simplified cost models to represent the travel time or distance, such as Manhattan distance or single-agent shortest path cost, and robots' coordination is often assumed to be optimistic, for example, ignoring the collision avoidance problem. 
This assumption does not hold in MAPD problems, where congestion and path conflicts substantially affect task completion cost.

\subsection{MAPD Approaches}
% \subsubsection{Offline Approaches}
Several works address MAPD from an offline perspective, which assumes all tasks are known and the computation time is given upfront. 
In \citet{nguyen2019generalized_mapd_asp}, the authors start to generalise the combined problem of path finding and task assignment and solve it with answer set programming. They proposed a three-phase method, which scales to only 20 agents.
\citet{liu2019task} propose a two-stage approach that first models the task assignment problem as a TSP problem, and then computes the task sequences for each agent and then plans execution paths using MAPF techniques. 
In \citet{honig2018conflict_cbs_ta}, authors introduce CBS-TA, which combines optimal task assignment with CBS-based path planning and solves this problem optimally. \citet{lam2025optimal_bcp_mapd} also propose BCP-MAPD that uses branch-and-cut-and-price to solve the combined problem optimally offline.
These methods provide strong solution quality but do not support online execution, and they scale poorly beyond hundreds of agents due to their combinatorial complexity.

% \subsubsection{Online Approaches}
In contrast, online MAPD methods handle dynamically generated tasks during execution. Token passing (TP) and Token Passing with Task Swaps (TPTS) \cite{ma2017lifelong_token_passing} are two decentralised methods that solve MAPD in two stages. In TP, agents select the closest task and plan their path to it one by one, and once a task is assigned to an agent, the assignment becomes fixed and cannot be swapped. While TPTS allows swapping for tasks that have not been picked up yet. \citet{ma2017lifelong_token_passing} also proposed a centralised algorithm, CENTRAL, which uses the Hungarian Method to solve the task assignment problem and computes paths using CBS \cite{sharon2015conflict}.
RMCA \cite{chen2021integrated} is another online method that integrates task assignment and path planning and solves them at the same time. RMCA starts with an initial assignment and plan for each agent, and then improves the solution within the runtime available.
These methods can handle an online environment, but they often only compute suboptimal solutions for fewer than hundreds of agents due to their computational complexity. 

\section{Traffic-Guided Planner for Path Planning}
In our framework, we decouple task assignment from path planning and solve the task assignment problem. For path planning, we directly use a recent state-of-the-art online MAPF planner, Guided PIBT \cite{chen2024traffic_planner}. This planner operates efficiently in large-scale online MAPF settings by reasoning about future congestion and using \textit{guide paths} as heuristics for determining next actions. We also integrate its traffic-aware cost models and guide paths into our task assignment model (described in later sections).

\subsubsection{Traffic-Aware Cost Models}
The planner first plans a time-independent guide path for each agent using focal search. During the search, two types of congestion are estimated and incorporated into edge costs. When traversing an edge $e$ from vertex $v_1$ to $v_2$, the planner considers:
%\begin{itemize}
\\ \textbf{Vertex Congestion ($p_{v}$)}
%    \item \textbf{Vertex Congestion ($p_{v}$)}: 
This estimates the total delay that will occur in the future if an agent enters vertex $v$, calculated using 
    $p_{v} = \left\lceil \frac{n_v - 1}{2} \right\rceil$
    where $n_v$ is the total number of agents entering vertex $v$ on its planned time-independent path.
%    \item \textbf{Contraflow Congestion ($c_e$)}: 
\\ \textbf{Contraflow Congestion ($c_e$)}
This estimates the potentially large delays caused by agents being pushed by other agents to avoid collisions which intend to traverse $e$ in different directions. This is computed as
    $c_e = f_{v_1,v_2} \cdot f_{v_2,v_1}$
    where $f_{v_i,v_j}$ denotes the number of agents currently planned to traverse edge $e$ in the direction from $v_i$ to $v_j$.   
%\end{itemize}
% \\ \textbf{Future Traffic}

The planner then combines these values into edge cost 
$FCost(e) = 1 + p_{v_2} + c_e$.
% Time-independent paths are computed to minimise these costs. 
% The edge usages are updated incrementally after each agent is planned or replanned.

\subsubsection{Guide Path Planning and Refinement}
% Initially, the planner plans time-independent guide paths for each agent one by one that minimise the traffic cost (sum of edge weights along the path) while considering the congestion costs of the planned agents. 
After all initial paths are computed, the system refines them iteratively by replanning a subset of agents based on updated congestion estimates.
%and updates the edge weights accordingly within the given runtime. 
A rule-based solver, PIBT \cite{okumura2022priority_pibt}, then uses the resulting guide path as a heuristic to determine the movement for each agent while avoiding collisions.

% The key idea of this planner is to maintain a spatial \emph{guide path} for each agent to avoid congestion and use a single-step rule-based planner \cite{okumura2022priority_pibt} to try to follow the guide path as much as possible. At each timestep, the planner commits only the next move for each agent while continuously improving the guide paths within the remaining planning time.

% \yz{I might need to write more details for the integration.}
% This approach scales well to large instances, and we integrate it into our framework in two ways, first, the traffic cost produced by the planner can be used to update edge weights in the flow network. Second, we use the output of our spatial flow algorithm as initial guide paths for the planner. This kind of loose coupling helps: (i) the flow algorithm to compute the assignments which reflect more on real congestions; and (ii) the planner to warm-start quickly with the flow guide path within the tight planning time windows.
%The traffic estimations represent the weight on each edge of the map, which are calculated from the waiting time estimations of the current guide paths.

\section{Flow Network for Task Assignment}

In this section, we describe how we model the task assignment subproblem in online MAPD as a flow over a graph of the map. We begin with a baseline bipartite linear assignment formulation and then introduce our spatial flow-based approach, which provides better scalability and integrates naturally with path planning. We then discuss how it integrates with the planner. Finally, we analyse the time complexity of both methods. 

\begin{figure}[t]
\centering
\begin{minipage}{0.29\linewidth}
    \centering
    \includegraphics[width=\linewidth]{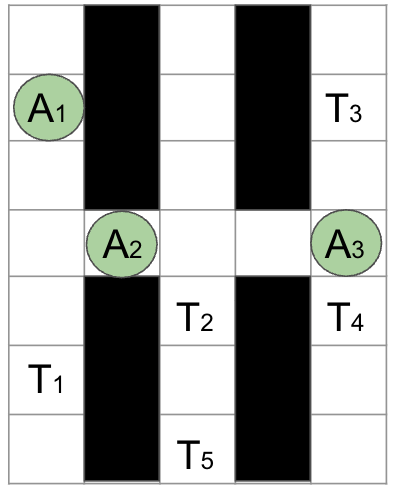}
    \caption*{(a)}
\end{minipage}
\hfill
\begin{minipage}{0.19\linewidth}
    \centering
    \includegraphics[width=\linewidth]{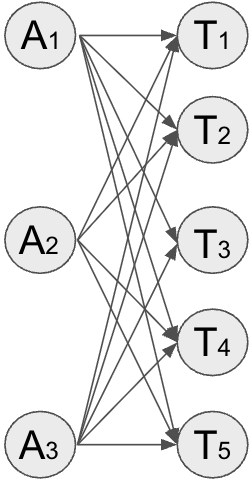}
    \caption*{(b)}
\end{minipage}
\hfill
\begin{minipage}{0.48\linewidth}
    \centering
    \includegraphics[width=\linewidth]{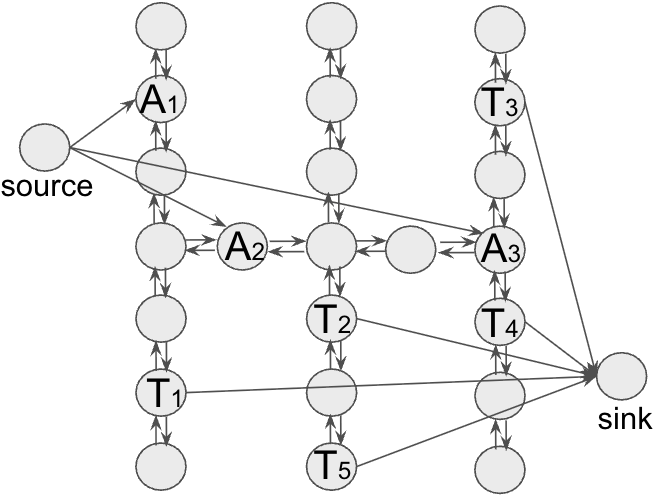}
    \caption*{(c)}
\end{minipage}

\caption{Illustration of task assignment models. (a): the example instance, where $A_i$ represents the agent, and $T_i$ represents the task (pickup location). (b): the bipartite linear assignment model, where each agent connects to every task and the edge cost is the shortest path distance. (c): the flow model, where the map is embedded directly as a flow network and the edge cost is the unit cost.}
\label{fig:assignment-comparison}
\end{figure}

\subsection{Linear Assignment Formulation}
A straightforward approach to task assignment is to model it as a minimum-cost bipartite assignment problem between available agents (agents are not delivering items) and available tasks (tasks are not picked up yet). 
An example of this formulation is shown in Figure \ref{fig:assignment-comparison}(b). 
For each agent-task pair, the edge cost is typically the shortest path distance between the agent's start location and the task location.
This is typically computed by Dijkstra, in which the search starts from the agent's start location and terminates when all tasks are reached.
Then, a complete bipartite graph is constructed from all the agent-task pairs.
The goal is to find a minimum-cost one-to-one assignment that minimises the total assignment cost. 
This problem can be solved optimally as a minimum-cost flow problem on a bipartite graph.

While this formulation is intuitive and provides optimal one-shot assignments, it has several limitations. First, it requires computing all agent-task distances, which becomes costly as the number of agents and tasks increases.\footnote{These can be precomputed if fixed unit costs are used, but not for traffic costs.} Second, the scalability of this formulation is limited in large-scale settings due to the quadratic number of edges. For example, $n=10000$ agents and $m=15000$ tasks will result in 150 million edges, and solving the problem with this model at this scale becomes challenging. 

\subsection{Flow-Based Model}
\paragraph{Graph Construction}

To overcome these limitations, we propose a spatial flow-based formulation that operates directly on $G$.
As shown in Figure \ref{fig:assignment-comparison}(c),
instead of computing agent-task costs explicitly, we embed both agents and tasks into a single flow network constructed from the map topology.
\begin{itemize}
    \item Each cell $v \in V$ of the grid map becomes a node in the directed graph. Then add edges between adjacent free cells.
    \item Add a dummy source node with edges to the current positions of agents that are not delivering, each with max flow one unit. % Allow at most one unit of flow from the source to each agent position.
    Force a flow from the source of $\min(m,n)$.
    \item Add a dummy sink node and edges from available pickup locations, each with max flow of one unit. 
    %Allow at most one unit to flow from each pickup location to the sink node.
    \item Edge costs that connect the map cells can be unitary (1) or an estimated cost, such as traffic estimations.
\end{itemize}
Unlike time-expanded formulations, we do not model time explicitly or enforce capacity constraints to prevent collisions in the flow network. Instead, we allow multiple units of flow to traverse the same edge, and leave the collision avoidance to path planners.

\paragraph{Assignment and Guide Path Retrieval from Flow}

This formulation enables the assignment to be solved as a single minimum-cost flow problem, from which we extract task assignments and guide paths for each agent by tracing the unit flow from the agent's current location to a task node in the flow network. 
The computed flow indicates not only which agent should go to which task, but also suggests a spatial path toward the task, which is aware of traffic congestion and can be used as a guide path for the planners.
As shown in Algorithm \ref{alg:path_retrieval}, for each agent, we start from its corresponding location node and follow the outgoing edges with positive flow.
On each node, we select one outgoing edge with positive flow to reach the next node and subtract one unit of flow from the edge.
We continue traversing the graph until we reach a task node (i.e., a node with an outgoing edge to the sink),  and collecting the visited nodes along the way as the agent's guide path. Once a task node is reached, we assign the corresponding task to the agent and store the constructed path for the downstream planner. 
Note that Algorithm \ref{alg:path_retrieval} returns one of possibly many optimal assignments that can be derived from the input flow.

\begin{algorithm}[t]
\caption{Retrieval from Flow Solution}
\label{alg:path_retrieval}
\begin{algorithmic}[1]
\STATE \textbf{Input:} Directed flow network $G=(V, A)$ with flow values $f: A \rightarrow N$; set of agents $A$ 
\STATE \textbf{Output:} Assigned task and guide path for each agent
\FORALL{$a_i \in A$}
    \STATE $v \leftarrow$ node at agent $a_i$'s current location
    \STATE $P \leftarrow$ empty list
    \WHILE{$v$ is not a task node}
        \STATE Append $v$ to $P$
        \STATE Choose $e = (v,v')$ an edge with positive flow $f(e) > 0$  from current node $v$
        \STATE $f(e) \leftarrow f(e) - 1$. Remove one unit of flow from $e$
        \STATE $v \leftarrow v'$
    \ENDWHILE
    \STATE Assign task at $v$ to $a_i$
    \STATE Store $P$ as the guide path for $a_i$
\ENDFOR
\end{algorithmic}
\end{algorithm}

\noindent
\paragraph{Alternative Edge Costs}

The default edge cost for our flow model is the edge cost from the map, i.e., unit cost for grid maps. We also support other cost functions that incorporate different considerations of the problem. 
For example, using estimated traffic congestion cost helps agents avoid waiting at frequently blocked or delayed areas. Here, we present two alternative dynamic edge costs based on the traffic.
%\begin{itemize}
%    \item \textbf{Traffic Cost from Planner Estimations: } 
\\ \textbf{Traffic Cost from Planner Estimations:}
Here we use the same traffic costs as the planner \cite{chen2024traffic_planner}, which uses future traffic estimations to plan paths for agents.
We use edge costs $Fcost()$ based on the same traffic estimations in our flow model. That is, at each planning cycle, we use the guide paths of agents that are currently delivering items (computed by the planner) to compute $Fcost(e)$ for each edge in the flow model. Note that agents delivering items will not be reassigned a new task, so this estimate is stable.
%\item \textbf{Avg Waiting Time from Execution: }
\\ \textbf{Avg Waiting Time from Execution: }
To support integration with other planners that do not have traffic estimation, we propose an alternative edge cost model that calculates the average waiting time up to this point of the execution (past traffic). 
We maintain a record of agent waiting times on each edge during execution and incorporate these traffic statistics into the edge cost. For each directed edge $e \in E$, we track two values: the total waiting time $W_e$ to traverse $e$ and the total number of traversals $N_e$ of $e$.
The cost of $e$ is set to its historical average traversal time:
\begin{align*}
PCost(e) = 1 + 
\begin{cases} 
\frac{W_e}{N_e} \text{ if $N_e > 0$},\\
0 \text{ otherwise}.
\end{cases}
\end{align*}
which is the unit cost plus the average wait time.
Additionally, we apply a decay factor $\gamma \in (0,1]$ to both $W_e$ and $N_e$ at each planning window.
We update $W_e$ and $N_e$ at each step using:
%We say an agent waits $t$ to traverse an edge $e (v_i, v_j)$ if the current actions are using $e$, and the previous actions before reaching $v_j$ are $t$ times of waitings on $v_i$: 
\begin{align*}
W_e & \leftarrow \gamma W_e + 
\begin{cases} 
$t$ \text{ if an agent traverses $e$ after waiting $t$},\\
0 \text{ otherwise}.
\end{cases}\\
N_e & \leftarrow \gamma  N_e + 
\begin{cases} 
1 \text{ if an agent traverses $e$ after waiting $t$},\\
0 \text{ otherwise}.
\end{cases}
\end{align*}
This helps the model emphasize more recent congestion observations while discounting older traffic conditions.
%\end{itemize}
%

\subsubsection{Planner Integration}
We integrate our flow model with the planner in two ways. 
\emph{(1) Traffic Cost Estimation from Planner to Flow:}
As also illustrated in previous subsections, we model the edge cost in flow using the edge cost estimates based on agents' guide paths and expected future traffic from the planner ($FCost$). This ensures the flow solver uses the same heuristic as the planner, and assigns tasks in a way that avoids regions likely to become congested. As a result, task assignments and subsequent path planning are optimised with respect to the same underlying traffic model.
%    \item \textbf{Guide Path Initialisation from Flow to Planner}: 
\emph{(2) Guide Path Initialisation from Flow to Planner:} 
    Although the planner is fast and scalable, initialising guide paths for thousands of agents on large maps can exceed the strict planning time limit (e.g., one second). Therefore, we warm-start the planner by using the path extracted from the flow solution.
    Once a minimum-cost flow is computed, we extract guide paths for each agent directly from the solution (as described in Algorithm~\ref{alg:path_retrieval}). These paths indicate not only the task assigned to each agent but also an initial route toward the task. We then pass these guide paths to the planner as initial guide path. This accelerates path generation by avoiding redundant search and helps to start the refinement process more quickly.
%\end{itemize}

% This loose coupling between task assignment and path planning improves overall system efficiency. It enables the flow solver to incorporate planner-informed cost models, while also allowing the planner to leverage optimised guide paths from assignment—yielding better coordination with minimal overhead.

\subsection{Complexity Analysis}
%The flow-based approach scales well as the number of agents and tasks increases and does not suffer from the bottleneck of edge weights computation.
We analyse and compare the computational complexity of our flow-based task assignment model against the baseline assignment problem.
Throughout the analysis, we use $n \leq |V|$ and $m \leq |V|$ and the fact that on a graph of a grid map with $|V|$ nodes and $|E|$ edges, we have $|E|=\mathcal{O}(|V|)$.
Note that, because the intention is to use edge costs that may not be integer, solving techniques that rely on \emph{cost scaling} \cite[Chapter 10]{ahuja1993network} are generally not appropriate for either approach.

\paragraph{Linear Assignment}
In the assignment approach, there are two stages of computation:
\begin{itemize}
    \item Edge weights computation: for each agent, we perform a shortest path search (e.g., Dijkstra) on the map of $|V|$ nodes and $|E|$ edges: $\mathcal{O}(n (|V| + |E|)\log |V|)) = \mathcal{O}(n |V|\log |V|)).$
    \item Solving an unbalanced assignment problem \cite{ramshaw2012unbalanced_assignment} takes, in the worst case: $\mathcal{O}(\min(n^2,m^2) m)$
    time, using the fact that, if $n \leq m$, only one of the closest $n$ tasks to an agent can be assigned to that agent, hence only $n^2$ edges are necessary, and, similarly, $m^2$ edges if $m \leq n$.
    % $$\mathcal{O}(n^2 \log n + nm)$$
\end{itemize}
Thus, the total time complexity of this approach is
$\mathcal{O}(n|V|(m+\log|V|)).$
%When $n$ is large (e.g., tens of thousands of agents), this becomes computationally expensive both in time and memory, especially due to the $n \times m$ cost matrix.

\paragraph{Network Flow}
Our flow-based model avoids explicit agent-task distance computation and operates on a simple graph representation of the map. The network is constructed directly from the map:
\begin{itemize}
    \item Nodes: $|V| + 2 = \mathcal{O}(|V|)$, including one node for each map cell plus a source node and a sink node.
    \item edges: $|E| + n + m = \mathcal{O}(|V|)$, including one edge for each traversable map edge plus $n$ edges that connects the source node to $n$ agents' current positions and $m$ edges that connects the sink node to $m$ task locations.
\end{itemize}
Using the algorithm given by \citet{orlin1993min_cost_flow}, given at most $|V|$ edges are capacitated, we find $\mathcal{O}(|V|^2\log^2 |V|).$
The Network Simplex, which we use in our experiments, has the same worst-case time complexity, supposing edge costs do not exceed a constant \cite{tarjan1997network_flow_analysis}.

%The resulting min-cost flow problem is solved using the Network Simplex algorithm, which runs in $\mathcal{O}(U \cdot m \cdot \log n)$ in practice, where $U$ is the maximum edge capacity and $m$ is the number of edges. In our model, the maximum edge capacity is up to $n$. As a result, the practical runtime is:
%$$\mathcal{O}( n \cdot (|E| + n + m) \cdot \log |V|)$$
%Typically, since agent paths are distributed across the map, the effective flow per edge is low, and Network Simplex runs in near-linear time. For this reason, a more practical estimate of the runtime is: 
%$$\mathcal{O}((|E| + n + m) \cdot \log |V|)$$
Unlike linear assignment, the spatial flow model scales primarily with the map size rather than the number of agent-task pairs. 
Indeed, if the number $n$ of agents grows linearly with $|V|$ (e.g. 50\% agent density), then the worst-case time complexity of solving the linear assignment is $\mathcal{O}(|V|^3)$, whereas that of the Network Flow is still $\mathcal{O}(|V|^2\log^2 |V|)$.
This makes it more suitable for large-scale MAPD problems, where $n \times m$ can easily exceed $|E|$ and $|V|$ by orders of magnitude.

\section{Experiments}
We implement the whole framework in C++ on top of the traffic planner \cite{chen2024traffic_planner}. For the linear assignment and network flow solver, we use the open-source library LEMON \cite{dezsHo2011lemon}, which, on these instances, was faster than competitors such as Gurobi.
The experiments are conducted on a cloud instance with 32GB RAM, 16 AMD EPYC-Rome CPUs.
We run one map from the standard grid-based Multi-Agent Path Finding (MAPF) benchmarks \cite{sturtevant2012benchmarks} and two warehouse-type maps with distance-biased distributions from a recent MAPF/MAPD competition called League of Robot Runners (LoRR) \cite{chan2024league}. The maps are:
\begin{itemize}
    \item \textit{Random (R)}: a 64 $\times$ 64 map with 3270 traversable cells and 20$\%$ random generated obstacles. The agent team size are tested from 400 to 2000, increasing by 400. 
    \item \textit{Warehouse Small (WS)}: a 33 $\times$ 57 map with 1277 traversable cells. Among the free cells, there are 40 ``E'' locations that represents the working stations in the warehouse, and 342 ``S'' locations that presents the items locations that needs to be picked up. 
    %The maps represents a typical MAPD scenario of fulfillment center.
    The agent team size are tested from 200 to 600, increasing by 100. 
    \item \textit{Sortation Large (SL)}: a 140 $\times$ 500 map with 54320 traversable cells. There are 620 ``E'' locations and 31540 ``S'' locations. The agent team size are tested from 4000 to 20000, increased by 4000. 
    % \item \textit{Warehouse Large (WL)}: a 140 $\times$ 500 map with 38586 traversable cells. There are 352 ``E'' locations and 25249 ``S'' locations. The maps represents a typical MAPD scenario of mail sortation center.
    % The agent team size are tested from 4000 to 16000, increased by 4000. 
    %\yz{I think I might need to add some figures for how the warehouse maps look like}
\end{itemize}
Note the range of agents is deliberately chosen so that we get to the point of having too many agents on the map, and hence throughput reduces. 
We keep the number of tasks in the task pool to 1.5 times the number of agents. For the task distributions, the tasks for \textit{Random} are sampled randomly, and the tasks for the other maps are selected only from ``E'' and ``S'' locations and are generated based on a distance-based warehouse distribution model used in LoRR, where tasks with fewer distance to working stations will have higher probability to be selected. 
%\pjs{If this is the model used by Robot runners then say it!}

\subsection{Experiment 1: Runtime of different methods}

\begin{table*}[t]
\centering
\small
\begin{tabular}{|c|c|c c|cccc|}
\hline
\multirow{2}{*}{\textbf{Map}} & \multirow{2}{*}{\textbf{$n$}} & \multicolumn{2}{c|}{\textbf{No Timeout (\% vs Greedy)}} & \multicolumn{4}{c|}{\textbf{With 1s Timeout (\% vs Greedy)}} \\
\cline{3-8}
& & \textbf{Greedy*} & \textbf{Flow-Unit Cost*} & \textbf{Greedy} & \textbf{Flow-Unit Cost} & \textbf{Flow-Traffic} & \textbf{Flow-Avg Waiting} \\
\hline
R  & 400   & 6936 & \textbf{6980 (↑0.63\%)} & 6925 & 6972 (↑0.68\%) & 6964 (↑0.56\%) & \textbf{6975 (↑0.72\%)} \\
R  & 800   & 12688 & \textbf{12871 (↑1.44\%)} & 12660 & \textbf{12826 (↑1.31\%)} & 12745 (↑0.67\%) & 12787 (↑1.00\%) \\
R  & 1200  & 15839 & \textbf{16211 (↑2.35\%)} & 15874 & 16331 (↑2.88\%) & 16205 (↑2.09\%) & \textbf{16402 (↑3.33\%)} \\
R  & 1600  & 16172 & \textbf{16971 (↑4.94\%)} & 16316 & 16383 (↑0.41\%) & \textbf{17097 (↑4.79\%)} & 17001 (↑4.20\%) \\
R  & 2000  & 15411 & \textbf{15626 (↑1.40\%)} & 15265 & 16101 (↑5.48\%) & 15939 (↑4.42\%) & \textbf{16111 (↑5.54\%)} \\
\hline
WS & 200   & 3508 & \textbf{3639 (↑3.73\%)} & 3512 & \textbf{3642 (↑3.70\%)} & 3596 (↑2.39\%) & 3632 (↑3.42\%) \\
WS & 300   & 4773 & \textbf{4954 (↑3.79\%)} & 4779 & \textbf{4919 (↑2.93\%)} & 4863 (↑1.76\%) & 4816 (↑0.77\%) \\
WS & 400   & 5570 & \textbf{5673 (↑1.85\%)} & 5494 & 5718 (↑4.08\%) & \textbf{5761 (↑4.86\%)} & 5637 (↑2.60\%) \\
WS & 500   & 5791 & \textbf{5900 (↑1.88\%)} & 5894 & 5829 (↓1.10\%) & \textbf{6174 (↑4.76\%)} & 5819 (↓1.27\%) \\
WS & 600   & 5585 & \textbf{5836 (↑4.49\%)} & 5593 & 5652 (↑1.05\%) & \textbf{6185 (↑10.59\%)} & 5678 (↑1.52\%) \\
\hline
SL & 4000  & 13776 & \textbf{14490 (↑5.18\%)} & 13642 & 13173 (↓3.44\%) & 13449 (↓1.41\%) & \textbf{14094 (↑3.32\%)} \\
SL & 8000  & 26355 & \textbf{27614 (↑4.78\%)} & 25831 & 18507 (↓28.37\%) & 25569 (↓1.01\%) & \textbf{26177 (↑1.34\%)} \\
SL & 12000 & 33952 & \textbf{35912 (↑5.77\%)} & 34414 & 23136 (↓32.81\%) & 33355 (↓1.72\%) & \textbf{34568 (↑0.45\%)} \\
SL & 16000 & 31428 & \textbf{33960 (↑8.05\%)} & 31993 & 26490 (↓17.21\%) & \textbf{34277 (↑7.14\%)} & 33975 (↑6.19\%) \\
SL & 20000 & 29029 & \textbf{31852 (↑11.42\%)} & 27323 & 24501 (↓10.33\%) & \textbf{31658 (↑15.91\%)} & 29477 (↑7.09\%) \\
% \hline
% WL & 4000  & 13088 & \textbf{13544 (↑3.48\%)} & 12968 & 13484 (↑3.98\%) & 13082 (↑0.88\%) & \textbf{13613 (↑4.97\%)} \\
% WL & 8000  & 23754 & \textbf{24857 (↑4.65\%)} & \textbf{23605} & 22961 (↓2.73\%) & 22689 (↓3.88\%) & 23197 (↓1.73\%) \\
% WL & 12000 & 23386 & \textbf{25211 (↑7.80\%)} & 23098 & 23332 (↑1.01\%) & \textbf{24325 (↑5.31\%)} & 23746 (↑2.8\%) \\
% WL & 16000 & 19603 & \textbf{22084 (↑12.63\%)} & 19273 & 18922 (↓1.82\%) & \textbf{21010 (↑9.01\%)} & 19568 (↑1.53\%) \\
\hline
\end{tabular}
\caption{
Throughput results across different maps and team sizes. 
%Throughput is defined as the number of tasks completed over 1000 timesteps.
Column 1 is the map name, column 2 is the team size (number of agents) and column 2-7 show the throughput of different methods.
Methods marked with ``*'' are unconstrained (no timeout) for task assignment and initial guide path computation, followed by 1s refinement.
The remaining methods operate under a strict 1-second real-time limit per timestep. Percentages in parentheses indicate relative improvement over the respective \textbf{Greedy} baseline. Bold values highlight the best-performing method within each group.
}
\label{tab:throughput-table}
\end{table*}

We compare the runtime performance of \textbf{Linear Assignment} (treating the cost computation time as free), \textbf{Linear Assignment + Dijkstra} (including cost computation using Dijkstra), and \textbf{Flow} (with unit cost as the edge cost). 
The time limit for returning assignments and actions is set to 10 minutes, and we simulate for 1000 timesteps.
We use \textit{Warehouse Small} and \textit{Sortation Large} to illustrate runtime behaviour across scales. 

\begin{figure}[t]
\centering
\begin{minipage}{0.91\columnwidth}
    \centering
    \includegraphics[width=\linewidth]{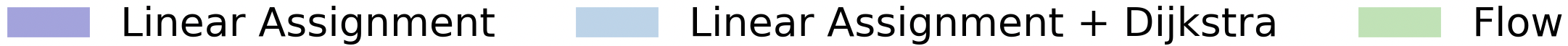}
    % \caption*{(a) Warehouse Small}
\end{minipage}
\begin{minipage}{0.49\columnwidth}
    \centering
    \includegraphics[width=\linewidth]{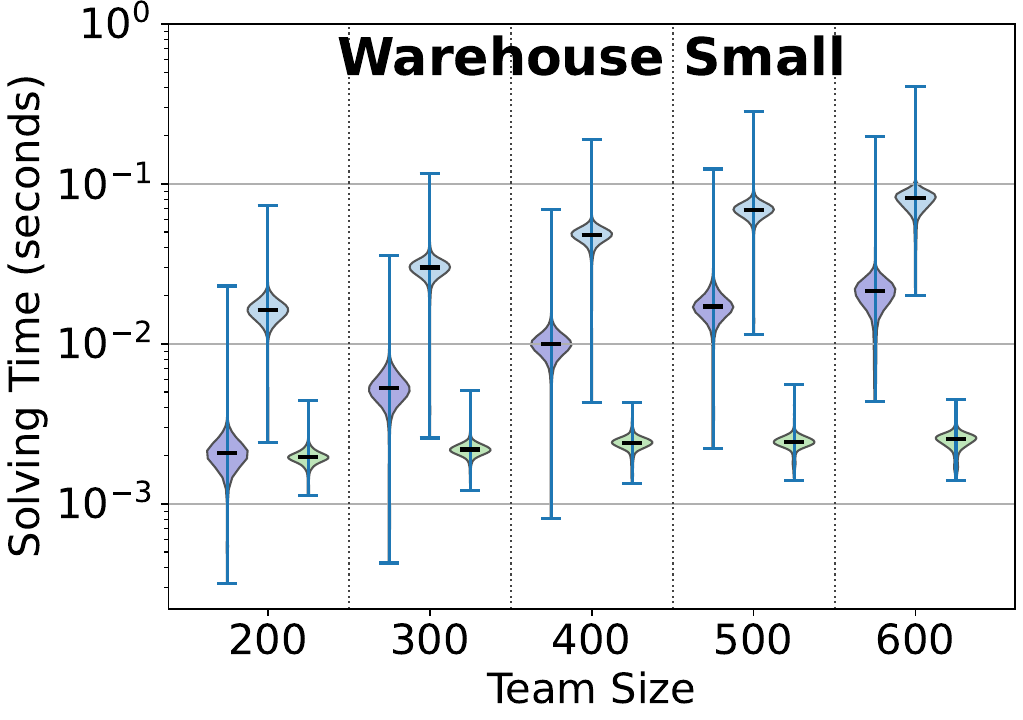}
    % \caption*{(a) Warehouse Small}
\end{minipage}
\hfill
\begin{minipage}{0.48\columnwidth}
    \centering
    \includegraphics[width=\linewidth]{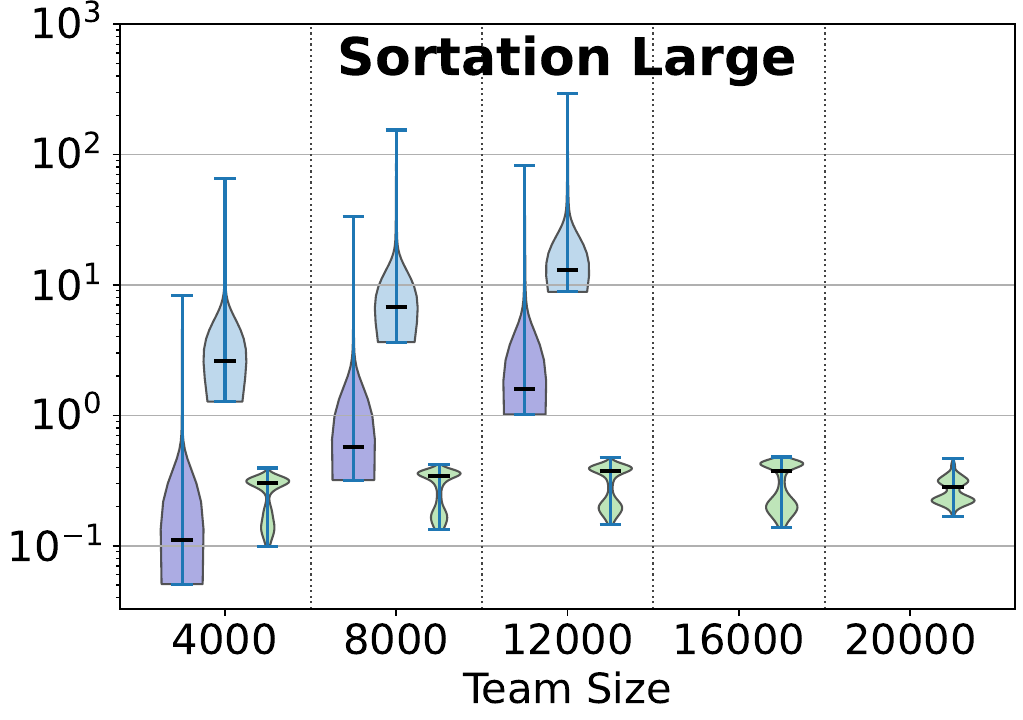}
    % \caption*{(b) Sortation Large}
\end{minipage}
\caption{Solving time distributions for different methods and team sizes on two maps. For team sizes 16000 and 20000 on Sortation Large, Linear Assignment fails to compute even the first step within 10 minutes.}
\label{fig:runtime-comparison}
\end{figure}

Figure \ref{fig:runtime-comparison} shows the runtime distributions of different methods across different timesteps. Note that during executions, the time for Linear Assignment varies during execution, because the number of available the number of agents and tasks varies.
%Since during executions, the active number of tasks and agents are smaller than the initial phase, we also show the runtime for the first timestep in Table \ref{tab:runtime_first}.
Overall, Flow achieves consistently low solving times across all team sizes. On \textit{Warehouse Small}, flow has a consistent runtime under 0.01s regardless of the team size, while linear assignment and the edge computation both show increasing runtime as team size grows. The performance gap is more dramatic on \textit{Sortation Large}. Linear assignment and its edge cost computation have a quadratic runtime growth and even fail to complete within 10 minutes at team sizes 16000 and 20000. In contrast, Flow maintains low solving times, scaling efficiently even at this scale.

\subsection{Experiment 2: Throughput of different methods}
We now evaluate the throughput performance of different task assignment methods, including our flow-based models and greedy, under varying team sizes and map structures.
Throughput is defined as the number of tasks completed over a 1000-timestep simulation. 
For greedy, we refer to greedily assigning task to agents according to the shortest path distance, the distance is assumed to be computed and cached on demand, and shared by the planner. Greedy only assigns new free agents with tasks and does not allow task swapping, which is the best greedy version we found in our experiments.\footnote{Note greedy is very fast, hence not compared in Experiment 1.} Table~\ref{tab:throughput-table} summarises the results. The methods are grouped into two categories.

% Table~\ref{tab:throughput-table} summarises the results. The methods are grouped into two categories.

\subsubsection{Unconstrained Time Limit} Columns 3-4 (Greedy* and Flow-Unit Cost*) correspond to the setting where task assignment and initial guide path computation are unconstrained in each timestep, and the path refinement has a 1s timeout limit. These methods benefit from sufficient time to compute good path plans, which reveals the potential of each strategy when computation is not a bottleneck.

In this setting, we observe that flow-based assignment consistently outperforms Greedy, even with a unit cost (gains from 0.6\% to over 11\%). The benefits are particularly apparent in large-scale warehouse settings. 
This illustrates the value of global optimisation for task-to-agent assignment, particularly when the global optimisation does not have too much time overhead, since the flow-based method is fast. 
However, as team size increases, the challenge shifts from task assignment to path planning. In particular, we observe that the initial guide path computation becomes the primary bottleneck, i.e., taking up to 26 seconds for 20000 agents on Sortation Large. 

% \begin{figure}[t]
% \centering
% \begin{minipage}{0.49\linewidth}
% \centering
% \includegraphics[width=\linewidth]{fig/task_random.pdf}
% \end{minipage}
% \hfill
% \begin{minipage}{0.49\linewidth}
% \centering
% \includegraphics[width=\linewidth]{fig/task_sortation.pdf}
% \end{minipage}
% \caption{Percentage of assignable agents and tasks over time on Random (2000 agents) and Sortation Large (20000 agents). ``Assignable'' means the number of agents/tasks that can be assigned or swapped.}
% \label{fig:assignable-agents}
% \end{figure}

% Another key finding is that the importance of task assignment varies significantly across map types and task distributions. In random maps with randomly sampled task distributions, many agents are able to pick up nearby tasks just one step away. As a result, the assignment problem in such settings is trivial, and throughput is dominated by local decisions. In contrast, warehouses and sortation centres impose spatial separation between pickup and delivery points, which makes task assignment a more critical component of overall performance.
%As shown in Figure \ref{fig:assignable-agents}, nearly all agents are in delivering and only very few are assignable. In contrast, warehouses and sortation centres impose spatial separation between pickup and delivery points, which makes task assignment a more critical component of overall performance.

\subsubsection{Strict Real-time Constraints}
Columns 5–8 represent the strict real-time MAPD setting, where each timestep is limited to a total of 1s time limit for assignment and path planning. For our flow-based method, we test three edge costs: \textbf{Unit Cost}, \textbf{Traffic} (Traffic Cost from Planner Estimations) and \textbf{Avg Waiting} (Avg Waiting Time from Execution). For Avg Waiting, we set $\gamma = 0.9$.
We only warm-start the planner with the flow solution when the cost model is \textit{Traffic} or \textit{Avg Waiting}, because the flow solution with unit cost is simply the shortest path, which does not avoid any congestion.

Overall, through the available time becomes less, the flow-based model outperforms greedy in most cases. Flow-Unit Cost performs well in smaller or less congested environments; the two congestion-aware variants, \textbf{Flow-Traffic} and \textbf{Flow-Edge Waiting}, consistently outperform both Greedy and unit-cost flow in congested scenarios. In particular, in situations like 16000-20000 agents in Sortation Large and 400-600 agents in Warehouse Small, Flow-Traffic under a strict timeout limit has better throughput than that with no timeout. In addition, Flow-Avg Waiting also shows similar performances to traffic estimations, which demonstrates that this estimation is suitable when used with planners that does not produce traffic estimation. 

We also observe a significant decrease for Flow-Unit Cost on Sortation Large. This is mainly because of the overhead for computing or recomputing (due to frequent task swaps) the guide path in the planner. 
For example, across 1000 timesteps, Flow-Unit Cost fails to find an initial solution within 1s for 997 times on Sortation Large with 20000 agents. 
Since we currently reschedule tasks every timestep, this indicates that less frequent scheduling may be needed in large-scale scenarios.
The other two flow-based variants, which can produce the initial guide path to the planners, are less affected by these runtime overheads.
Note Greedy does not suffer from this issue, because it does not allow task swapping, hance there is no need to replan guide paths.

\subsection{Experiment 3: Flow vs existing MAPD approaches}
We compare our approach against RMCA \cite{chen2021integrated}, the existing state-of-the-art method for online MAPD. We test our method on their task release policy and problem instances. In this setting, $f \in {2,5,10}$ number of tasks are released per timestep, and we measure the makespan (timesteps to finish total tasks) for 500 tasks and $n \in {50,80,100}$ number of agents. For a given $f$ and $n$, we test 25 instances.
We count the number of steps that exceed a 1s runtime limit for RMCA as the total number of timeouts.

Table~\ref{tab:makespan-runtime} reports the average makespan for our method and RMCA and the average number of timesteps for RMCA (with standard deviations).
Our flow-based approach consistently achieves lower makespan across all instances. The performance gap becomes larger under high task densities (e.g., $f = 10$), where RMCA incurs many timeouts due to its higher computational overhead. In contrast, our method maintains consistent planning times ($<1s$) across all timesteps.
\begin{table}[t]
\centering
\small
\renewcommand{\tabcolsep}{5pt}
\begin{tabular}{|c|c|cc|c|}
\hline
\multirow{2}{*}{\textbf{$f$}} & \multirow{2}{*}{\textbf{$n$}} & \multicolumn{2}{c|}{\textbf{Avg Makespan}} & \textbf{Avg Timeouts} \\[0.3ex]
\cline{3-4}
& & \textbf{RMCA} & \textbf{Flow} & \textbf{RMCA} \\
\hline
2 & 50 & 324.96 $\pm$ 6.99 & \textbf{299.24} $\pm$ 7.80 & 74.88 $\pm$ 6.97 \\
2 & 80 & 288.00 $\pm$ 3.91 & \textbf{242.00} $\pm$ 8.99 & 38.00 $\pm$ 3.91\\
2 & 100 & 287.20 $\pm$ 3.7 & \textbf{227.20} $\pm$ 12.01 & 37.20 $\pm$ 3.70  \\
5 & 50 & 319.72 $\pm$ 7.58 & \textbf{271.64} $\pm$ 8.28 & 219.68 $\pm$ 7.55 \\
5 & 80 & 218.48 $\pm$ 5.65 & \textbf{197.88} $\pm$ 5.10 & 118.48 $\pm$ 5.65  \\
5 & 100 & 185.52 $\pm$ 4.65 & \textbf{172.92} $\pm$ 6.51 & 85.48 $\pm$ 4.60 \\
10 & 50 & 315.48 $\pm$ 7.85 & \textbf{271.16} $\pm$ 8.48 & 268.36 $\pm$ 9.57  \\
10 & 80 & 216.24 $\pm$ 4.84 & \textbf{192.88} $\pm$ 5.06 & 175.52 $\pm$ 7.09  \\
10 & 100 & 184.12 $\pm$ 4.97 & \textbf{166.36} $\pm$ 7.30 & 147.04 $\pm$ 6.50 \\
\hline
\end{tabular}
\caption{Average makespan (column 3-4) between RMCA and Flow and average number of timeouts for RMCA (column 5) under varying team sizes $n$ and task frequencies $f$. }
\label{tab:makespan-runtime}
\end{table}

\begin{table}[h]
\centering
\small
\renewcommand{\tabcolsep}{5pt}
\begin{tabular}{|c|c|c|cc|c|}
\hline
\multirow{2}{*}{\textbf{Map}} & \multirow{2}{*}{\textbf{$n$}} & \multirow{2}{*}{\textbf{Simulate}}  & \multicolumn{2}{c|}{\textbf{Throughput}} & \textbf{Avg Time(s)} \\[0.5ex]
\cline{4-5} \cline{5-5}
& & \textbf{Steps} & \textbf{Greedy} & \textbf{Flow} & \textbf{Flow} \\
\hline
Orz & 10000 & 2000 & 1031 & \textbf{1042} & 0.367($\pm0.02$) \\
Orz & 20000  &2000 & 2031 & \textbf{2139} & 4.138($\pm7.24$) \\
IH & 10000 & 5000 & 2624 & \textbf{2646} & 19.819($\pm1.20$) \\
IH & 20000 & 5000 &5698 & \textbf{5902} & 18.201($\pm1.56$) \\
\hline
\end{tabular}
\caption{Throughput between Greedy and Flow-based methods and solving time for Flow. We simulate 2000 steps for Orz and 5000 steps for IH. 
%\yz{It's too few timesteps compared to the map size, not sure if we should include, but at least the resutls are okay.}
}
\label{tab:large-throughput}
\end{table}

\subsection{Experiment 4: Flow on ultra-large maps}
We further evaluate the scalability of our flow-based assignment framework on two ultra-large maps: \textbf{Orz900d (Orz)}, size 1491 $\times$ 656 with 96603 free cells from \citet{sturtevant2012benchmarks}, and \textbf{mp\_2p\_01 Iron harvest (IH)}, 1912 $\times$ 1800 with 6545639 free cells from \citet{harabor2022benchmarks_ih}. We simulate with different team sizes $n \in {10000,20000}$ with Greedy and Flow-Unit Cost. 
In such massive environments, there is no practical way to compute or store accurate heuristics (e.g., all-pairs shortest paths). Thus, we use a lightweight setup: Manhattan distance as a heuristic and PIBT \cite{okumura2022priority_pibt}, a simple rule-based solver, for path planning. To reduce the overhead of frequent assignment computation at this scale, we schedule task assignment using flow every $k$ timesteps. That is, the flow solver is invoked once every $k$ steps, and the resulting assignments are fixed for the window. If an agent finishes its task during this interval, it remains unassigned until the next round. We set $k = 10$ for Orz and $k = 30$ for IH.

As shown in Table~\ref{tab:large-throughput}, the flow-based method achieves higher throughput while maintaining tractable runtime. On IH, which has over 6 million free cells, the flow solver remains stable with runtime under 20 seconds, even for 20000 agents.
An additional benefit of the flow model is that, our flow-based model produces globally optimal task assignments with respect to true shortest-path distances without explicitly computing or storing any pairwise paths, because the minimum-cost flow is solved over the grid itself.

% from the standard benchmark set \cite{sturtevant2012benchmarks}. 
% \begin{itemize}
%     \item \textbf{Orz900d (Orz)}: a 1491 \times 656 grid with 96603 free cells.
% \end{itemize}

\section{Conclusion}
In this work, we study the task assignment in online MAPD. Our primary contribution is a spatial flow-based task assignment framework that directly operates on the map and avoids costly pairwise distance computations. The model supports real-time execution, planner integration, and high scalability. We further explore two light-weight congestion-aware edge costs to demonstrate the flexibility of the flow model. These models serve as initial examples of how traffic information can be embedded into the flow to improve coordination.
Experiments show that it outperforms baselines in large-scale settings, especially when combined with planner warm-starts and congestion estimates.
Future work includes designing more accurate and adaptive edge cost models, improving the flow network structure, and extending the framework to support other MAPD variants.

% \begin{figure*}
%     \centering
%     \includegraphics[width=\linewidth]{fig/warehouse_large_eg.jpg}
%     \caption{Caption}
%     \label{fig:enter-label}
% \end{figure*}

\bibliography{flow}

\end{document}